\documentclass[letterpaper]{article} 
\usepackage[]{aaai2026}
\usepackage{times}  
\usepackage{helvet}  
\usepackage{courier}  
\usepackage[hyphens]{url}  
\usepackage{graphicx} 
\def\UrlFont{\rm}  
\usepackage{natbib}  
\usepackage{caption} 
\usepackage{algorithm}
\usepackage{algorithmic}
\usepackage{booktabs}
\usepackage{multirow}
\usepackage{makecell}
\usepackage{amsmath}
\usepackage{amssymb}
\usepackage{amsfonts} 
\usepackage{appendix}
\usepackage{newfloat}
\usepackage{listings}
\usepackage{xcolor}
\DeclareCaptionStyle{ruled}{labelfont=normalfont,labelsep=colon,strut=off} 
\floatstyle{ruled}
\newfloat{listing}{tb}{lst}{}
\floatname{listing}{Listing}
\nocopyright 

\title{From Scores to Skills: A Cognitive Diagnosis Framework for Evaluating Financial Large Language Models}

\author{
    Ziyan Kuang\textsuperscript{\rm 1,2},
    Feiyu Zhu\textsuperscript{\rm 1,2},
    Maowei Jiang\textsuperscript{\rm 3},
    Yanzhao Lai\textsuperscript{\rm 4},
    Zelin Wang\textsuperscript{\rm 1,2},
    Zhitong Wang\textsuperscript{\rm 5},
    Meikang Qiu\textsuperscript{\rm 6},
    Jiajia Huang\textsuperscript{\rm 3},
    Min Peng\textsuperscript{\rm 1,2},
    Qianqian Xie\textsuperscript{\rm 1,2}\thanks{Corresponding author. Email: xieq@whu.edu.cn},
    Sophia Ananiadou\textsuperscript{\rm 7} 
}

\affiliations {
    \textsuperscript{\rm 1}School of Artifical Intelligence, Wuhan University\\
    \textsuperscript{\rm 2}Center for Language and Information Research
, Wuhan University\\
    \textsuperscript{\rm 3}School of Computer Science, Nanjing Audit University\\
    \textsuperscript{\rm 4}Southwest Jiaotong University\\
    \textsuperscript{\rm 5}Beijing University of Financial Technology\\
    \textsuperscript{\rm 6}Computer and Cyber Sciences, Augusta University\\
    \textsuperscript{\rm 7}Computer Science, University of Manchester\\
}

\usepackage{bibentry}

\begin{document}

\maketitle

\begin{abstract}
Large Language Models (LLMs) have shown promise for financial applications, yet their suitability for this high-stakes domain remains largely unproven due to inadequacies in existing benchmarks.
Existing benchmarks solely rely on score-level evaluation, summarizing performance with a single score that obscures the nuanced understanding of what models truly know and their precise limitations. They also rely on datasets that cover only a narrow subset of financial concepts, while overlooking other essentials for real-world applications. 
To address these gaps, we introduce FinCDM, the first cognitive diagnosis evaluation framework tailored for financial LLMs, enabling the evaluation of LLMs at the knowledge-skill level, identifying what financial skills and knowledge they have or lack based on their response patterns across skill-tagged tasks, rather than a single aggregated number.
We construct CPA-KQA, the first cognitively informed financial evaluation dataset derived from the Certified Public Accountant (CPA) examination, with comprehensive coverage of real-world accounting and financial skills. It is rigorously annotated by domain experts, who author, validate, and annotate questions with high inter-annotator agreement and fine-grained knowledge labels.
Our extensive experiments on 30 proprietary, open-source, and domain-specific LLMs show that FinCDM reveals hidden knowledge gaps, identifies under-tested areas such as tax and regulatory reasoning overlooked by traditional benchmarks, and uncovers behavioral clusters among models.
FinCDM introduces a new paradigm for financial LLM evaluation by enabling interpretable, skill-aware diagnosis that supports more trustworthy and targeted model development, and all datasets and evaluation scripts will be publicly released to support further research.\footnote{https://github.com/WHUNextGen/FinCDM}
\end{abstract}


\section{Introduction}
Large language models (LLMs) are increasingly applied to financial tasks~\cite{nie2024survey}, but their suitability for such high-stakes domains remains largely untested~\cite{xie2024finben}. 
While numerous open-domain benchmarks such as MMLU~\cite{hendrycks2021measuring}, HELM~\cite{liang2022holistic}, and HLB~\cite{duan2024hlb} have demonstrated the excellent general capabilities of LLMs like GPT-4~\cite{achiam2023gpt}, DeepSeek~\cite{liu2024deepseek}, and LLaMA~\cite{touvron2023llama}, their effectiveness in the financial domain-specific tasks remains largely unknown.

To address this uncertainty, several benchmarks~\cite{xie2023pixiu,xie2024finben,li2024investorbench,peng2025multifinben}, have been recently proposed to systematically evaluate the capabilities of LLMs specifically within the financial domain. 
However, existing financial LLM benchmarks~\cite{xie2024finben,li2024investorbench,peng2025multifinben} often fail to reflect what matters for those tasks and applications, namely what the model knows, what it can reliably do, and where it is likely to fail.
First, this issue stems from \textbf{score flattening}, where a dataset in existing benchmarks is reduced to a single number, making it unclear what knowledge the model has actually mastered.
For example, in FinQA from MultiFinBen \cite{peng2025multifinben}, GPT‑4o, a general-domain LLM, and FinMA~\cite{xie2023pixiu}, a financial LLM, achieve similar overall accuracy scores. However, GPT‑4o tends to perform better on numerical computation tasks such as calculating the ``net change in cash'', while FinMA is stronger at handling finance-specific conceptual questions, such as identifying ``total equity''.
Second, many prior datasets also exhibit \textbf{coverage imbalance}, where examples disproportionately rely on certain types of financial knowledge or skills, making it difficult to evaluate how models perform across the full range of real-world requirements. For example, in the accounting questions of FinEval~\cite{guo2025fineval}, most items focus on a narrow set of concepts such as total revenue and net income, while overlooking other areas like equity changes or tax-related components.
\begin{figure*}[htp]
    \centering
    \includegraphics[width=1\linewidth]{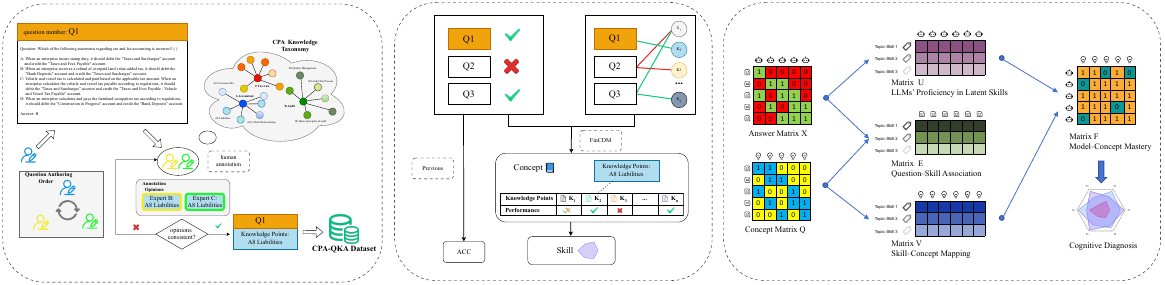}
    \caption{The whole framework of FinCDM.}
    \label{fig:framework}
\end{figure*}
To address these limitations, we introduce FinCDM, the first cognitive diagnosis evaluation framework for financial LLMs. Inspired by cognitive diagnosis model (CDM) in educational assessment, FinCDM evaluates LLMs like human examinees, identifying what financial skills and knowledge they have or lack based on their response patterns across skill-tagged tasks, rather than a single aggregated number.
First, we apply the non-negative matrix
co-factorization based CDM model to estimate LLM's mastery on different financial skills 
with collected responses of LLMs, addressing the problem of score flattening.  CDMs assume that the model's answer to a task reflects whether the model possesses all the underlying skills required by the question.
Therefore, by decomposing the responses to all questions based on knowledge labels annotated for each task, we can assess the model's proficiency in different knowledge, revealing where it excels and where it struggles.
Second, we introduce the first Chinese dataset named \texttt{CPA-QKA} rigorously annotated by domain experts, specifically constructed to support the knowledge-skill level evaluation of FinCDM, and provide comprehensive evaluation coverage. 
The dataset is grounded in the Certified Public Accountant (CPA) exam, the most recognized professional certification in accounting and financial reporting. 
We build the financial knowledge framework covering 70 key concepts from the content and skill specification outline of the CPA exam, ensuring comprehensive coverage of real-world financial concepts.
Under the framework, our dataset construction involves a rigorous, two-phase annotation process. Domain experts first author multiple questions for per concept, and then perform verification to ensure accuracy, consistency, and high inter-annotator agreement (IAA). 
These verified questions are annotated with precise financial knowledge and skill tags, also employing IAA-based quality control.
By annotating existing benchmarks using our framework, we further highlight their limitations, illustrating their narrow coverage of financial concepts and underscoring the need for the broader and deeper skill assessment that FinCDM provides.
Figure~\ref{fig:framework} illustrates the complete workflow of FinCDM.

Based on FinCDM, we evaluate 30 representative LLMs, including proprietary models, open-source general-purpose models of varying sizes, and finance-specific models. Our results support three key findings. First, FinCDM reveals significant differences in models’ mastery of financial knowledge that are not captured by traditional aggregate metrics. For example, Doubao shows strong performance on Chinese-specific regulations and specialized accounting areas, while Gemini demonstrates superior understanding of ``Debt Restructuring'', ``Lease'', ``Post-Balance Sheet Events'', showcasing robust mastery in general accounting concepts. Despite these differences, both models achieve similar overall scores on FinQA, highlighting how traditional aggregate metrics can obscure underlying capability variation. Second, our evaluation illustrates that prior benchmarks predominantly assess a narrow subset of financial concepts, resulting in inadequate coverage. In contrast, our new dataset systematically reveals previously overlooked weaknesses in areas like deferred tax liabilities, lease classification, and regulatory ratios, which are concepts rarely tested in existing benchmarks but critical in real-world applications. Third, by analyzing the knowledge-skill level mastery profiles produced by FinCDM, we identify latent associations between financial concepts and reveal distinct clusters of models exhibiting similar skill acquisition patterns. For example, GPT-3.5 and DeepSeek-VL share strengths in financial reporting and valuation, while FinGPT and FinQwen exhibit aligned capabilities in regulation and macroeconomic reasoning, reflecting different domain specialization strategies.

In summary, our major contributions are:
\begin{enumerate}
    \item We propose FinCDM, the first cognitive diagnosis framework for financial LLM evaluation, which moves beyond aggregate metrics by assessing models’ proficiency at the knowledge-skill level.
    \item We construct a new dataset and structured financial knowledge framework derived from the CPA exam, with high-quality human annotations. This supports reliable evaluation of FinCDM and exposes the narrow coverage of existing benchmarks.
    \item We apply FinCDM to a broad set of 30 proprietary, open-source, and finance-specific models, uncovering their knowledge gaps, strengths, and behavioral patterns, offering actionable insights for model development and deployment.
\end{enumerate}

\section{Related Work}
\subsection{Financial Benchmark}
Numerous benchmarks have been developed to evaluate LLMs in financial domains, covering tasks such as extraction, QA, reasoning, simulation, and retrieval. Early efforts like PIXIU ~\cite{xie2023pixiu} and FinBen ~\cite{xie2024finben} offer broad coverage but rely on aggregate metrics. Benchmarks like FinanceBench ~\cite{islam2023financebench}, BizBench ~\cite{krumdick2024bizbench}, and BizFinBench ~\cite{lu2025bizfinbench} focus on QA and business reasoning, while multilingual and low-resource evaluation is addressed by CFinBench ~\cite{nie2024cfinbench}, CFLUE ~\cite{zhu2024benchmarking}, FinEval ~\cite{zhang2023fineval}, Golden Touchstone ~\cite{wu2024golden}, Plutus-ben ~\cite{peng2025plutus}, and the more recent MultiFinBen \cite{peng2025multifinben}. 
FinMTEB \cite{tang2025finmteb} cover embedding-based retrieval and classification. Structured reasoning is evaluated in FinDABench ~\cite{liu2024findabench}, Fino1 ~\cite{qian2025fino1}, FinChain ~\cite{xie2025finchain}, and agentic decision-making in InvestorBench ~\cite{li2024investorbench}, AlphaFin ~\cite{li2024alphafin}, and AveniBench ~\cite{klimaszewski2025avenibench}. 
Multimodal understanding is addressed by M$^3$FinMeeting ~\cite{zhu2025m}and FinAudio ~\cite{cao2025finaudio}.FinTagging ~\cite{wang2025fintagging} and FinDER ~\cite{choi2025finder} focus on fine-grained financial concept extraction and retrieval-augmented QA from 10-K filings and XBRL reports. Despite their breadth, these benchmarks rely on task-level aggregate metrics and lack concept-aware diagnostics.

\subsection{Financial Dataset Design}
A wide range of data sets have been developed to evaluate financial LLM in reasoning, QA, information extraction, summarization, and multimodal tasks. Numerical reasoning datasets include FinQA~\cite{chen2021finqa}, TAT-QA~\cite{zhu2021tat}, ConvFinQA~\cite{chen2022convfinqa}, DocFinQA~\cite{reddy2024docfinqa}, and FinanceQA~\cite{mateega2025financeqa}, which require multistep or conversational numerical inference over financial reports. FinTextQA~\cite{chen2024fintextqa}, FinLLMs~\cite{yuan2024finllms}, and FinTruthQA~\cite{xu2024fintruthqa} focus on long-form or formula-based QA, while SEC-QA~\cite{lai2024sec} and MultiHiertt~\cite{zhao2022multihiertt} introduce multidocument and table-text hybrid challenges. Structured extraction is addressed by FinTagging~\cite{wang2025fintagging}, FiNER-ORD~\cite{shah2023finer}, FiNER-139~\cite{loukas2022finer}, FinRED~\cite{sharma2022finred}, REFinD~\cite{kaur2023refind}, which allow evaluation of concept and relationship levels aligned with financial taxonomies. For numeral understanding, FinNum~\cite{chen2019overview}, FinNum-2~\cite{chen2020overview}, and FiNCAT~\cite{ghosh2022fincat} annotate fine-grained semantic types or numeral claims in finance text. The summarization is covered by ECTSum~\cite{mukherjee2022ectsum}, FNS-2020~\cite{el2020financial}, focusing on earnings calls and annual reports. Chinese and multilingual resources include FinEval~\cite{zhang2023fineval}, CFLUE~\cite{zhu2024benchmarking}, and UCFE~\cite{yang2024ucfe}, enabling QA and NLU in non-English contexts. Multimodal datasets such as FinMME~\cite{luo2025finmme} and FinLMM-R1~\cite{lan2025finlmm} support chart, image, and document–text alignment. Despite covering diverse formats and tasks, most existing datasets are task-driven with limited concept coverage.

\subsection{Cognitive Diagnosis Model}
Cognitive diagnosis models (CDMs), rooted in educational assessment, aim to infer mastery of latent knowledge attributes from observed response behaviors. Early interpretable models such as the Deterministic Inputs, Noisy 'And' gate model (DINA)\cite{de2009dina}, the Deterministic Inputs, Noisy 'Or' gate model (DINO)\cite{templin2006measurement}, and the Generalized DINA (GDINA) model\cite{de2011generalized} employ binary latent attributes and probabilistic response functions; while valuable for interpretation, they depend heavily on accurate Q matrix specification and strict parametric assumptions~\cite{gu2019sufficient}. To mitigate these limitations, matrix factorization-based approaches, such as MF‑DINA and logistic matrix factorization, embed examinees and attributes into low‑dimensional spaces, enabling more flexible and robust modeling of item-attribute interactions. Neural network-based CDMs such as NeuralCD~\cite{wang2022neuralcd} and recent graph-based models such as RCD~\cite{gao2021rcd} use deep architectures and disentangled graph learning to model nonlinear and noise-robust concept, exercise, and student representations. These neural approaches offer superior predictive power on large datasets but can suffer from overfitting and reduced interpretability when evaluated with few entities and many attributes.

\section{Method}
\subsection{Preliminaries}

In educational psychology, assessment plays a central role not only in certifying achievement but also in supporting learning and guiding instruction. Rather than reporting a single aggregate score, modern diagnostic assessments aim to provide interpretable and actionable insights into learners' mastery of specific knowledge components~\cite{frederiksen1993test, leighton2007cognitive}. This fine-grained feedback is foundational to formative assessment, enabling educators to design targeted interventions and fostering trust in the evaluation process~\cite{kuh2011piecing}.

Cognitive diagnosis models (CDMs)~\cite{leighton2007cognitive} were developed to meet these goals by explicitly modeling how a learner’s observed responses on assessment items reflect their their latent mastery of specific knowledge concepts. 
In the CDM framework, a particular knowledge domain is defined by $K$ latent knowledge components (or attributes), and each learner is associated with a binary mastery profile $\boldsymbol{\alpha} \in \{0,1\}^K$, where $\alpha_k = 1$ indicates mastery of the $k$-th skill.
Each assessment item $i \in \{1, \dots, I\}$ is linked to a subset of these skills through the expert-defined binary Q-matrix $\mathbf{Q} \in \{0,1\}^{I \times K}$, where $q_{ik} = 1$ signifies that answering item $i$ correctly requires mastery of skill $k$. Given a learner's binary response vector $\mathbf{X} \in \{0,1\}^I$, where $X_i = 1$ denotes a correct answer to item $i$, CDMs estimate the probability of correctly answering each item by modeling the relationship among the learner's mastery profile $\boldsymbol{\alpha}$, the item's skill requirements $\mathbf{q}_i$, and additional model-specific parameters $\boldsymbol{\theta}$:

\[
P(X_i = 1 \mid \boldsymbol{\alpha}, \mathbf{Q}, \boldsymbol{\theta}) = f_i(\boldsymbol{\alpha}, \mathbf{q}_i; \boldsymbol{\theta})
\]
where, $f_i$ denotes the item response function for item $i$.
The primary learning objective in CDMs is to accurately infer each learner's skill mastery profile $\boldsymbol{\alpha}$ and estimate the model parameters $\boldsymbol{\theta}$ from observed response data. 
By inferring fine-grained knowledge states, CDMs enable transparent and skill-specific evaluation, offering richer insights than traditional test scores.

\subsection{Dataset Curation}
\begin{figure}
    \centering
    \includegraphics[width=1\linewidth]{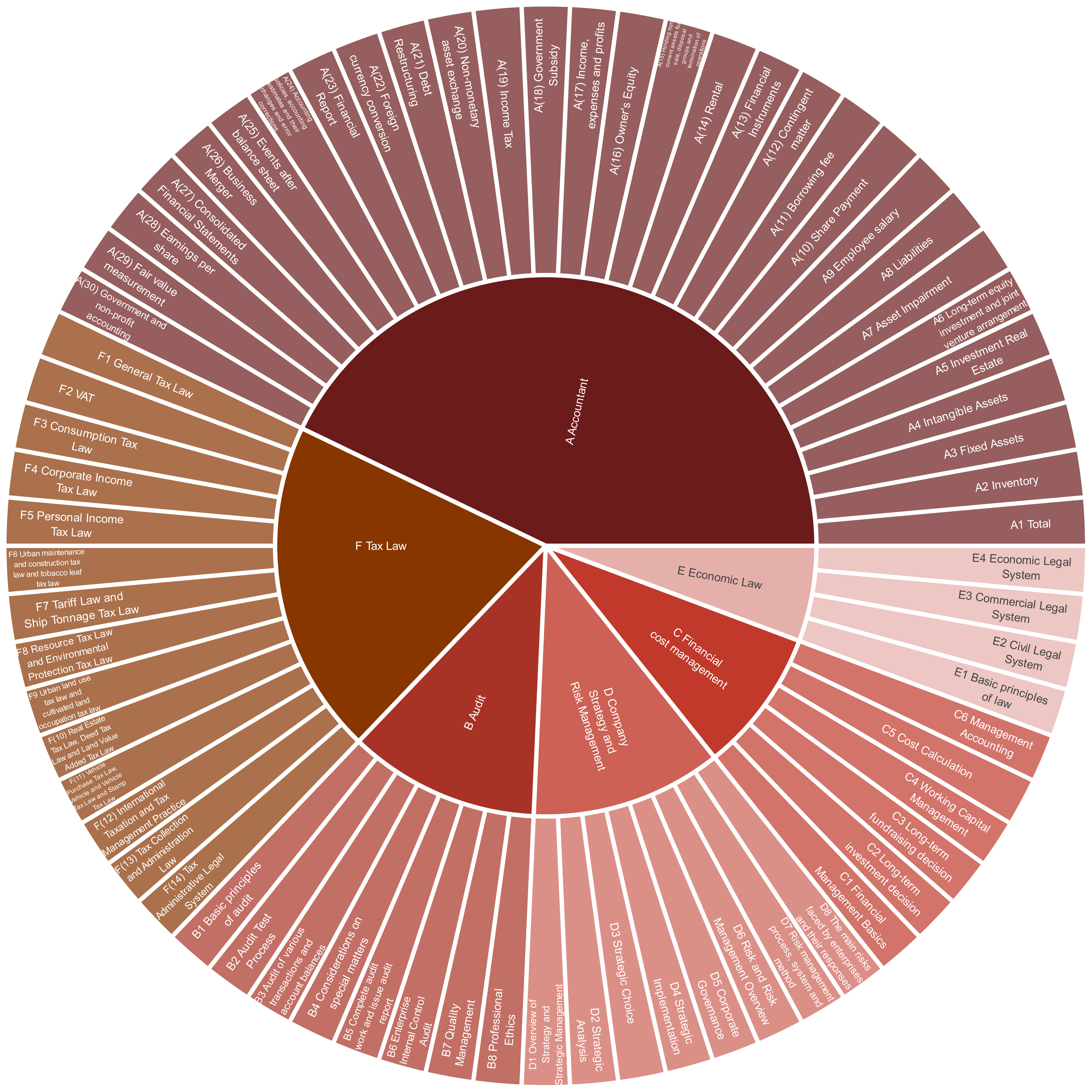}
    \caption{70 financial concepts covered by CPA-KQA.}
    \label{fig:concepts}
\end{figure}

Existing benchmarks typically rely on aggregate score evaluations and cover only a limited subset of financial concepts. To address these limitations, we propose a new dataset \textbf{CPA-KQA} with domain experts-authored financial questions, to support the knowledge-skill level evaluation of financial LLMs using CDMs, and provide comprehensive knowledge coverage. 
To construct CPA-KQA, we referenced the CPA examination, a highly recognized professional certification in the financial industry, to ensure a comprehensive coverage of financial knowledge.
It covers core financial areas including accounting, auditing, financial cost management, corporate strategy and risk management, economic law, and tax law.
We build the financial knowledge framework covering 70 core concepts, derived from the content and skill specification outline of the CPA exam \footnote{\url{https://www.cicpa.org.cn/xxfb/tzgg/202502/W020250228535255887805.pdf}}, such as “fixed assets”, “liabilities”, and “long-term investment decisions”, as shown in figure \ref{fig:concepts}. 

\textbf{Expert Annotation and Quality Validation:}
Our data annotation involves two phases including question authoring and financial concepts tagging. 
The annotation team comprised three domain experts: an undergraduate student, a master's student, and an associate professor specializing in finance. For each of the 70 financial concepts, three distinct questions were crafted by three annotators to ensure multiple cognitive perspectives, resulting in a total of 210 high-quality items.
Annotators were explicitly instructed to craft questions ensuring clarity and precision. 
To ensure reliability and consistency, we conducted rigorous quality validation. In the first phase, each authored question was independently reviewed by the two other experts (excluding the original author) to assess relevance, clarity, and alignment with the intended financial concept. Discrepancies were resolved through collaborative discussions. Inter-annotator agreement for question-concept mappings in this phase was measured using Krippendorff’s alpha.

In the second stage, all questions are independently annotated by the same two experts, who assign one or more knowledge labels selected from a predefined set of 70 financial concepts. If the annotators’ assigned knowledge concepts are consistent with those originally provided by the question setter, the question proceeds without modification. In cases where the annotators disagree with each other or with the question setter, a three-way discussion is held to resolve the disagreement and determine the final label. Annotation consistency in this stage is measured using the $\sigma$ score and the KS-based measure~\cite{braylan2022measuring}. 

As shown in Table \ref{tab:agreement_cpa}, our annotation achieved high consistency, with a Krippendorff’s alpha of 0.937, a $\sigma$ score of 0.9904, and a KS-based measure of 1. These results confirm the high reliability and quality of our CPA-KQA dataset. Further annotation details are provided in the appendix.

\begin{table}[t]
\centering
\begin{small}
\begin{tabular}{l|cc}
\toprule
&\multicolumn{2}{c}{\textbf{Dataset}}\\
\hline
\textbf{Metric} & \textbf{CPA-KQA}&\textbf{Fineval-KQA}\\
\midrule
Krippendorff's Alpha & 0.937 & -\\
\hline
$\sigma$ & 0.9904 & 0.9208 \\
KS-based measure & 1 & 1\\
\bottomrule
\end{tabular}
\end{small}
\caption{Inter-annotator agreement metrics for human expert annotations on CPA-KQA and Fineval-KQA.}
\label{tab:agreement_cpa}
\end{table}

\subsubsection{FinEval-KQA}
To further reveal the coverage of financial knowledge in existing datasets, we introduce the \textbf{FinEval-KQA} dataset, augmenting the FinEval question set~\cite{zhang2023fineval}, a widely used Chinese financial domain knowledge evaluation benchmark, with concept-level annotations based on the well-established framework of \textbf{CPA-KQA}. 
The FinEval dataset~\cite{zhang2023fineval} is a publicly available resource containing over 2,000 questions designed to evaluate financial models on a variety of financial knowledge areas. These questions span across multiple sub-domains, including financial analysis, accounting, and financial forecasting. We apply the CPA-KQA concept taxonomy to annotate a subset of FinEval with 101 questions that focuses on accounting-related knowledge. 

\textbf{Expert Annotation and Quality Validation:}
Our three expert annotators was tasked with mapping existing FinEval questions to the most relevant CPA-KQA concepts. A guideline document (see appendix for more details) was created to ensure that each question was annotated consistently with respect to the CPA-KQA taxonomy, and the annotators were instructed to focus on clarity, specificity, and relevance.
To ensure high-quality and consistent annotations, the inter-annotator agreement was evaluated using the same metrics as the original CPA-KQA annotations: $\sigma$ and KS-based measure. We conducted a reconciliation process in which annotators discussed any discrepancies in their annotations to ensure alignment. The results showed strong consistency, with $\sigma$ of 0.9208 and KS-based measure of 1, demonstrating the robustness of our annotation process.


\subsection{Evaluation Framework}
Building upon our annotated datasets, we now illustrate how we assess the mastery of financial knowledge of different LLMs. Specifically, using our newly constructed \textbf{CPA-KQA}, we can obtain the observed response data from LLMs, which forms our \textbf{response matrix} $X$, and leverage the expert-annotated \textbf{Q}.
We employ a non-negative matrix co-factorization inspired by SNMCF~\cite{yu2023snmcf} and DINA~\cite{de2009dina}, modeling the process of evaluating LLM skill mastery explicitly through a probabilistic generative model. 
Our framework assumes the following generative process:

\begin{enumerate}
    \item Latent skill representation for questions: For each question \( i \), generate a latent skill requirement vector \(\mathbf{e}_i\) from a Gamma prior:
    \[
    \mathbf{e}_i \sim \text{Gamma}(a, b), \quad \mathbf{e}_i \in \mathbb{R}_{\geq0}^{T}.
    \]

    \item Latent skill proficiency for models: For each LLM \( j \), generate a latent skill proficiency vector \(\mathbf{u}_j\) from a Gamma prior:
    \[
    \mathbf{u}_j \sim \text{Gamma}(c, d), \quad \mathbf{u}_j \in \mathbb{R}_{\geq0}^{T}.
    \]

    \item Latent skill to financial concept mapping: For each financial concept \( k \), generate a latent skill-to-concept association vector \(\mathbf{v}_k\) from a Gamma prior:
    \[
    \mathbf{v}_k \sim \text{Gamma}(e, f), \quad \mathbf{v}_k \in \mathbb{R}_{\geq0}^{T}.
    \]

    \item Generation of observed responses \( X \): Each observed binary response \( x_{ij} \) indicating whether LLM \( j \) correctly answers question \( i \) is drawn from a Bernoulli distribution parameterized by the alignment between the latent skill vectors:
    \[
    x_{ij} \sim \text{Bernoulli}\left(\sigma(\mathbf{e}_i^\top \mathbf{u}_j)\right), \quad \sigma(z) = \frac{1}{1 + e^{-z}}.
    \]

    \item Generation of question-concept association \( Q \): Similarly, the observed binary association \( q_{ik} \), indicating whether question \( i \) is related to concept \( k \), is drawn from a Bernoulli distribution parameterized by the alignment between their latent vectors:
    \[
    q_{ik} \sim \text{Bernoulli}\left(\sigma(\mathbf{e}_i^\top \mathbf{v}_k)\right), \quad \sigma(z) = \frac{1}{1 + e^{-z}}.
    \]
\end{enumerate}


Through this generative formulation, we aim to factorize the matrices 
\( X \) and \( Q\) into low-dimensional latent representations with non-negative constraints:
\[
X \approx E U, \quad Q \approx E V, \quad E, U, V \geq 0
\]
where \(E \in \mathbb{R}_{\ge0}^{M \times T} \) represents the relationship between questions and \( T \) latent skills, capturing how strongly each question is related to each latent skill. \( U \in \mathbb{R}_{\ge0}^{T \times N} \) captures the proficiency of each model in the latent skills. Each row in \( U \) corresponds to a latent skill, and each column represents the proficiency of model \( j \) in that skill. \( V \in \mathbb{R}_{\ge0}^{T \times K} \) represents the relationship between latent skills and financial concepts, mapping the latent skills to the relevant concepts for each question.
We estimate these latent matrices by optimizing the following joint objective:
\begin{scriptsize}
\[
\min_{E,U,V\geq 0} \|W \circ (X - EU)\|_F^2 + \beta\|Q - EV\|_F^2 + \lambda_E\|E\|_F^2 + \lambda_U\|U\|_F^2 + \lambda_V\|V\|_F^2,
\]
\end{scriptsize}
The factorized matrices \( E \), \( U \), and \( V \) provide a rich, interpretable structure that reveals the relationships between questions, concepts, and models' capabilities. See appendix for the optimization process to inference these latent matrices.
Finally, we explicitly estimate each LLM's mastery of financial concepts by combining the learned latent matrices \( U \) and \( V \):
\[
F = U^\top V, \quad F \in \mathbb{R}_{\geq0}^{N \times K},
\]
\( F \) provides detailed, interpretable diagnostics, indicating the proficiency level of each LLM across the financial concepts.

\subsection{Benchmarking}
Based on FinCDM, we conduct a benchmark study covering both closed-source and open-source models, including those specifically tailored for the financial domain. In total, our evaluation involves over 30 Chinese-capable LLMs. Detailed model information can be found in Appendix. We evaluate models in the following categories:

\begin{enumerate}
    \item \textbf{Closed-source general models:}  These include GPT-4, GPT-4o, and GPT-4o-mini ~\cite{achiam2023gpt}; Claude 3.5 Sonnet, and Claude 3.7 Sonnet ~\cite{anthropic2024claude3}; Gemini 1.5 Pro, Gemini 1.5 Flash, and Gemini 2.5 Pro Experimental ~\cite{team2024gemini}; Grok-3 ~\cite{xai2024grok3}; Doubao-1.5-Pro-256k, and Doubao-1.5-Pro-32k ~\cite{bytedance2024doubao15pro}.

    \item \textbf{Open-source general models:} These include Baichuan2-13B-Chat ~\cite{baichuan2023baichuan2}, ChatGLM3-6B ~\cite{chatglm2024chatglm3}, Falcon-7B ~\cite{almazrouei2023falcon}, GLM-4-32B-0414 and GLM-4-9B-0414 ~\cite{glm2024chatglm}; Qwen2-72B-Instruct, Qwen2.5-7B-Instruct, Qwen3-0.6B, and Qwen3-235B-A22b ~\cite{yang2025qwen3,team2024qwen2} DeepSeek-Chat and DeepSeek-V3-0324 ~\cite{bi2024deepseek,liu2024deepseek} and Hunyuan ~\cite{tencent2025hunyuanA13B}.
    
    \item \textbf{Financial domain models:} These include Finma-7b-Full ~\cite{xie2023pixiu}, CFGPT2-7B ~\cite{li2023cfgpt}. 
\end{enumerate}

For the evaluation, we prompt (see appendix for the prompt used in our experiments) each LLM to answer the questions in our datasets using a consistent and controlled setup. The decoding configuration is set with a temperature of 1.2 to promote diversity in generated responses, allowing models to explore varied responses under the same input. The maximum generation length is capped at 64 tokens to ensure concise answers, and each prompt instructs the model to return relevant and informative outputs.
Each model generates 10 responses per question. The final performance scores are computed by averaging over these 10 responses, providing a more reliable and robust estimate of model proficiency.
All models are evaluated using a unified script and configuration pipeline to ensure fairness and comparability across systems.

\section{Results}
\subsection{Main Results}
\begin{table}[htb]
\centering
\small
\caption{Results on CPA-KQA and Fineval-KQA. The ``Con'' column means concept, which represents the number of concepts for which the model achieved a mastery probability greater than 0.9. The data is sorted in descending order for ease of viewing. }
\label{tab:rank_cpa}
\begin{tabular}{lll|lll}
\toprule
\multicolumn{3}{c|}{\textbf{CPA-KQA}} & \multicolumn{3}{c}{\textbf{Fineval-KQA}} \\
\cmidrule(lr){1-3} \cmidrule(lr){4-6}
Con & Model & Acc & Con & Model & Acc \\
\midrule
\textbf{40/70} & \textbf{GLM4} & 0.63 & \textbf{13/38} & Gemini1.5 & 0.66 \\
39/70 & Claude3.7 & 0.77 & \textbf{13/38} & Gemini2.5pro & 0.87 \\
37/70 & Qwen3-235b & 0.73 & 11/38 & Claude3.7 & 0.76 \\
36/70 & Doubao32k & 0.82 & 11/38 & Doubao1.5pro & \textbf{0.88} \\
36/70 & Qwen-max & 0.76 & 11/38 & Qwen2.5-72b & 0.77 \\
34/70 & Doubao256k & \textbf{0.84} & 10/38 & Claude3.5 & 0.69\\
34/70 & Qwen2.5-72b & 0.76 & 8/38 & Gpt4 & 0.75 \\
33/70 & Claude3.5 & 0.74 & 8/38 & Qwen-max & 0.75 \\
33/70 & Gemini2.5pro & \textbf{0.84} & 8/38 & Qwen3-235b & 0.79 \\
31/70 & Gemini1.5 & 0.64 & 7/38 & DBRX & 0.42 \\
29/70 & Hunyuan & 0.64 & 7/38 & Gpt4o-mini & 0.48 \\
27/70 & Grok3 & 0.63 & 7/38 & Qwen2.5-7b & 0.68 \\
25/70 & Deepseek-v3 & 0.65 & 7/38 & Glm4 & 0.69 \\
25/70 & Gpt4 & 0.63 & 7/38 & Baichuan2 & 0.39 \\
24/70 & Deepseek-chat & 0.65 & 6/38 & Doubao256k & 0.85 \\
24/70 & Qwen2.5-7b & 0.62 & 6/38 & Gpt4o & 0.49 \\
24/70 & Glm4-9b & 0.56 & 6/38 & Llama2-70b & 0.47 \\
22/70 & Gemini1.5pro& 0.59 & 6/38 & Llama3.1 & 0.48 \\
21/70 & Gpt4o-mini & 0.50 & 6/38 & Glm4-32b & 0.67 \\
20/70 & Gpt4o & 0.52 & 6/38 & CFGPT2 & 0.48 \\
19/70 & Glm4-32b & 0.57 & 5/38 & Deepseek-chat & 0.74 \\
19/70 & CFGPT2 & 0.38 & 5/38 & Deepseek-v3 & 0.74 \\
17/70 & Chatglm3 & 0.34 & 5/38 & Glm4-9b & 0.56 \\
17/70 & Baichuan2 & 0.39 & 5/38 & chatGlm3 & 0.37 \\
13/70 & DBRX & 0.38 & 4/38 & Gemini1.5pro & 0.54 \\
13/70 & Llama2 & 0.45 & 4/38 & Hunyuan & 0.70 \\
11/70 & Llama3.1 & 0.57 & 3/38 & Grok3 & 0.60 \\
10/70 & Qwen3-0.6b & 0.24 & 3/38 & Qwen3-0.6b & 0.29 \\
9/70 & Finma7b & 0.19 & 2/38 & Falcon & 0.20 \\
0/70 & Falcon7b & 0.10 & 1/38 & Finma7b & 0.19 \\
\bottomrule
\end{tabular}
\end{table}

Table~\ref{tab:rank_cpa} and Figure~\ref{fig:heatmap_cpa-kqa} summarize the knowledge-skill level performance of various LLMs on our CPA-KQA dataset. Our key findings are as follows:

\begin{figure*}[htb]
\centering
\includegraphics[width=\textwidth]{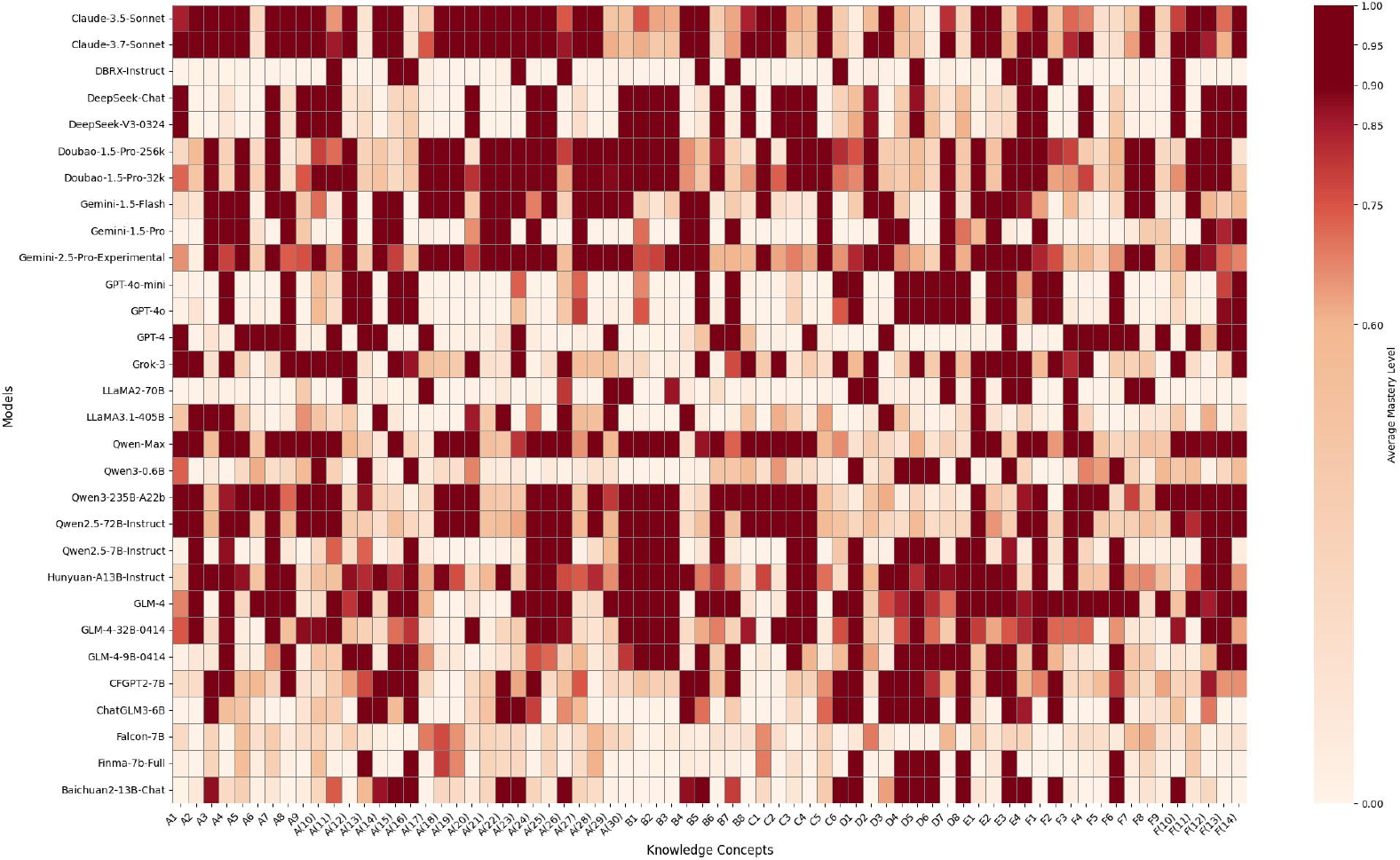}
\caption{Model knowledge mastery heatmap on the CPA-KQA dataset.}
\label{fig:heatmap_cpa-kqa}
\end{figure*}

\paragraph{The Role of FinCDM.}
FinCDM provides a more informative evaluation than aggregate metrics by uncovering how a model’s overall performance is distributed across individual financial concepts, highlighting both strengths and blind spots that remain hidden under traditional evaluation. As shown in Figure~\ref{fig:heatmap_cpa-kqa}, Claude 3.5 Sonnet achieves high average accuracy on CPA-KQA, but FinCDM reveals that this performance is concentrated in a narrow subset of concepts. These differences are not reflected in its aggregate score, which suggests uniformly strong performance. FinCDM exposes this imbalance by diagnosing both the breadth and depth of the model’s conceptual coverage, offering a more interpretable and actionable analysis of model capabilities.

\paragraph{Divergent Knowledge Specialization across High-Performing Models.}  
While many top-performing models achieve similar overall passing rates, our analysis further reveals substantial variation in the specific financial concepts each model masters. This divergence reflects differing strengths in subdomains of financial knowledge, suggesting that high aggregate accuracy does not necessarily imply uniform competence.
For instance, Gemini-2.5-Pro-Exp and Doubao-1.5-Pro-256k both attain high overall passing rates of 0.84, respectively. However, a closer inspection at the concept level reveals that Gemini excels in general accounting categories such as \textit{contingency} and \textit{Lease}, which are typically aligned with international financial reporting standards. In contrast, Doubao demonstrates stronger performance in financial cost management areas, particularly \textit{Long-term Investment Decisions}, \textit{Long-term Financing Decisions}, and \textit{Working Capital Management}, reflecting its expertise in financial management domains.
This pattern is consistent across other model pairs as well. 
These findings emphasize the importance of fine-grained, concept-level evaluation beyond aggregate metrics, and support the need for modular assessments that can reveal the unique strengths and limitations of each model.

\paragraph{Influence of Linguistic Resources on Model Performance.}  
Our evaluation highlights the substantial impact of linguistic resource availability on a model’s performance in domain-specific tasks. Models with limited Chinese language capabilities consistently underperform, both in aggregate accuracy and in concept-level mastery. For example, Falcon-7B, which lacks robust pretraining on Chinese corpora, achieves a passing rate of only 0.15 and demonstrates minimal competence across financial concepts.
These results underscore the necessity of adequate linguistic grounding for effective domain adaptation. Without sufficient coverage of the target language during pretraining or fine-tuning, models struggle not only with general comprehension but also with acquiring specialized financial knowledge.

\begin{figure}
    \centering
    \includegraphics[width=1\linewidth]{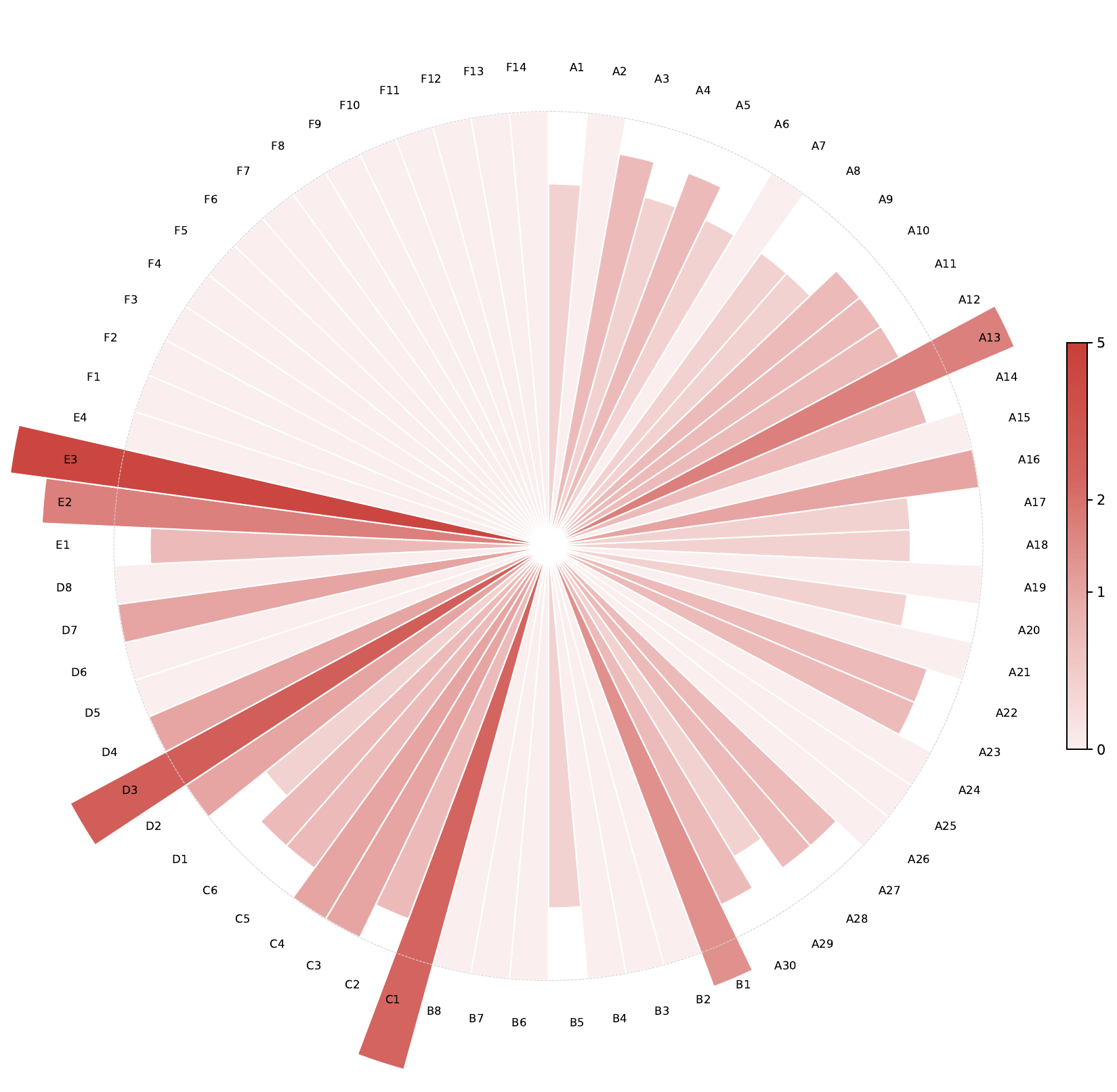}
    \caption {Structure of FinEval-KQA. The lighter-colored blocks represent concepts covered by FinEval-KQA, while the darker-colored blocks indicate concepts missing from it. The further a block extends from the center, the more frequently that concept was assessed.}
    \label{fig:fineval_structure}
\end{figure}

\paragraph{Evaluating Datasets on Knowledge Mastery.}  
To investigate how dataset structure shapes model understanding of financial knowledge, we evaluate multiple LLMs also on \textsc{FinEval-KQA}, an existing benchmark re-annotated at the knowledge-point level. As illustrated in Figure~\ref{fig:fineval_structure},
\textsc{FinEval-KPA} exhibits significant structural imbalance, with a majority of questions concentrated on a few specific financial concepts, notably \textit{Financial Instruments}, \textit{Fundamentals of Financial Management}, \textit{Strategic Choices}, \textit{Civil Law}, and \textit{Commercial Law}, the latter appearing as many as 13 times. This skewed distribution strongly influences model performance, causing mastery rates to be heavily dependent on the frequency and representation of concepts. In contrast, \textsc{CPA-KQA} maintains a more balanced representation across a broader range of financial concepts, enabling more robust and generalizable assessments of conceptual understanding.
For instance, Gemini1.5 demonstrates mastery across the relatively highest number of concepts but achieves a low overall accuracy due to limited mastery of frequently tested concepts such as extit{Commercial Law} , highlighting how uneven concept distribution can disproportionately impact model performance.

\subsection{Ablation Studies}
We further compare the performance of three categories of CDMs: Neural CDMs, Graph-based CDMs, and our employed Matrix Co-Factorization (MCF)-based method. Table~\ref{tab:ablation} presents the results in terms of accuracy (acc), area under the curve (AUC), and root mean square error (RMSE). 
To compute these metrics, we first use the inferred latent matrices \( E \) and \( U \) to reconstruct the predicted response matrix \( \hat{X} = EU \). We then compute the metrics by comparing these predictions against the observed response matrix \( X \).
MCF outperforms both baselines by a significant margin, achieving the highest accuracy and AUC, and the lowest RMSE. These results correspond to absolute gains of $+0.177$ in accuracy and $+0.146$ in AUC, and a reduction of $-0.167$ in RMSE compared to the strongest baseline, RCD. 

\begin{table}[H]
\small
\centering
\caption{Model performance comparison across architectures.}
\label{tab:ablation}
\resizebox{\columnwidth}{!}{%
\begin{tabular}{lccc}
\toprule
\textbf{Model} & \textbf{Accuracy (acc)} & \textbf{AUC} & \textbf{RMSE} \\
\midrule
Matrix Co-Factorization (MCF) & \textbf{0.9379} & \textbf{0.9873} & \textbf{0.2314} \\
Neural CDM (CNN-based)~\cite{wang2022neuralcd}          & 0.7140          & 0.7789          & 0.4394 \\
GNN-based CDM~\cite{gao2021rcd}               & {0.7469} & {0.8329} & {0.4110} \\
\bottomrule
\end{tabular}}
\end{table}

\subsection{Case Study}
To evaluate the reliability of our diagnostic framework, we conduct a focused case study on Claude 3.5, examining its mastery of two specific financial concepts: F3 and F5. According to our cognitive diagnosis analysis (Figure~\ref{fig:heatmap_cpa-kqa}), Claude 3.5 failed to demonstrate mastery in these two concepts. Upon closer inspection, we confirmed that the model answered all six related questions incorrectly (three questions per concept). This provides concrete evidence that our diagnostic framework effectively identifies specific knowledge gaps in financial LLMs.
To further verify the reliability of our framework, we enlisted five certified auditing experts to independently review the six incorrectly answered questions along with Claude 3.5’s responses. All five experts hold at least an undergraduate degree in finance and auditing. They were asked to annotate the primary financial concept tested by each question without being informed of the original concept labels.
As shown in Table~\ref{tab:annotation}, four of the five experts consistently assigned the questions to concepts F3 and F5. The fifth expert introduced minor variations, such as labeling one question as F2 or F4. The overall inter-annotator agreement was 0.80, indicating strong consensus among the experts.
These results demonstrate that our diagnosis framework aligns closely with expert judgment in identifying these conceptual deficiencies.


\begin{table}[H]
\small
\centering
\caption{Expert annotations of knowledge points associated with six incorrectly answered questions.}
\label{tab:annotation}
\resizebox{\columnwidth}{!}{%
\begin{tabular}{lcccccc}
\toprule
\textbf{Question Index} & 1 & 2 & 3 & 4 & 5 & 6 \\
\midrule
Expert 1 & F3 & F3 & F3 & F5 & F5 & F5 \\
Expert 2 & F3 & F3 & F3 & F5 & F5 & F5 \\
Expert 3 & F3 & F3 & F3 & F5 & F5 & F5 \\
Expert 4 & F2 & F3 & F4 & F5 & F5 & F5 \\
Expert 5 & F3 & F3 & F3 & F5 & F5 & F5 \\
\bottomrule
\end{tabular}}
\end{table}

\section{Conclusion}
This work presents FinCDM, the first cognitive diagnosis framework for evaluating financial LLMs beyond conventional aggregate metrics. Inspired by educational assessment, FinCDM diagnoses model proficiency at the knowledge-skill level, enabling interpretable insights into what financial concepts a model has mastered or misunderstood. To support this, we introduce CPA-KQA, a high-quality, expert-annotated benchmark grounded in the CPA exam, covering 70 core financial concepts with balanced knowledge representation. Through evaluations on 30 diverse LLMs, FinCDM reveals that models with similar overall scores often differ markedly in concept-level mastery, exposes coverage gaps in existing benchmarks, and uncovers latent specialization patterns across models. Future directions include multilingual extensions, incorporation of multimodal financial content, and leveraging diagnostic feedback to inform instruction tuning and benchmark design.

\bibliography{aaai2026}

\appendix
\section{Guidelines for Creating and Annotating CPA-KQA Items}\label{sec:cpa_annotation}
\subsection{Principles and Requirements for Item Creation}
\subsubsection{Consistency with CPA Exam Style}
Each assessment item must align closely with the CPA exam in terms of wording, question format, and option setting, reflecting the CPA exam's rigorous standards for professional knowledge.
\subsubsection{Originality and Accuracy}
All items must be original creations. Direct copying or quoting from existing CPA exam items or publicly available materials is strictly prohibited. Ensure accuracy, clarity, and absence of ambiguity in each item.
\subsubsection{Specificity and Clarity}
Each item should clearly assess only one specified financial knowledge point. Avoid incorporating content irrelevant to the specified knowledge point to ensure accurate and effective evaluation.
\subsubsection{Avoidance of Politically Sensitive Content}
Strictly avoid incorporating politically sensitive topics, events, or individuals in any assessment items. Content that may be viewed as politically controversial or sensitive is explicitly prohibited.
\subsubsection{Item Formats}
Create three items per knowledge point,  in Single-choice questions (at least one per knowledge point)
Provide four clearly distinguishable options (A, B, C, D), with exactly one correct answer.
\subsection{Required Format for Item Submission}
Submit each item strictly following this format:
\begin{itemize}
\item\textbf{Knowledge Point}: (Name of the Knowledge Point)
\item\textbf{Question Stem}: Clearly and completely stated
Options (only for single-choice items):
A. ...  B. ...  C. ...  D. ...
\item\textbf{Correct Answer}: (Option letter or detailed solution)
\end{itemize}
\subsection{Annotation and Quality Review Procedures}
Experts will create and annotate items as follows:
\begin{itemize}
    \item Each expert independently creates items for their assigned knowledge points.
    \item Upon completion, two other experts annotate each item on two dimensions:
    \textbf{Quality Annotation}: Mark as either "Usable" or "Unusable". Criteria for "Unusable" include unclear meaning, inaccurate wording, calculation errors, ambiguous answers, or controversial content.
    \textbf{Knowledge Point Consistency Annotation}: Verify if the item accurately corresponds to the specified knowledge point.
    \item Items marked "Unusable" or "Inconsistent" by both reviewers are deleted directly.
    \item Items marked by only one reviewer require one-time revision and must pass re-review.
    \item Deleted items must be replaced until each knowledge point has three approved items.
\end{itemize}

\section{Optimization Process for Non-negative Matrix
Co-factorization}\label{sec:cdm_optimization}
We estimate these latent matrices by optimizing the following joint objective:
\[
\begin{split}
\min_{E,U,V\geq 0} \,&\|W \circ (X - EU)\|_F^2 + \alpha\|Q - EV\|_F^2 \\
&+ \lambda_E\|E\|_F^2 + \lambda_U\|U\|_F^2 + \lambda_V\|V\|_F^2,
\end{split}
\]

with multiplicative update rules for convergence to a local optimum:
\[
E \leftarrow E \circ \frac{(W \circ X)U^\top + \alpha QV^\top}{(W \circ EU)U^\top + \alpha EVV^\top + \lambda_E E},
\]

\[
U \leftarrow U \circ \frac{E^\top(W \circ X)}{E^\top(W \circ EU) + \lambda_U U},
\]

\[
V \leftarrow V \circ \frac{\alpha E^\top Q}{\alpha E^\top EV + \lambda_V V},
\]
where the symbol \(\circ\) denotes element-wise multiplication, and division operations are also element-wise.

\begin{onecolumn}
\centering
\section{LLMs details}
\begin{table*}[htb]
\centering
\begin{tabular}{llllcc}
\toprule
Category & Model & Creator & Parameters & Access & Version Date \\
\midrule
\multirow{15}{*}{\centering Open-Source} & GLM-4-32B-0414 & Zhipu AI & undisclosed & API & 2025.4 \\
& GLM-4-9B-0414 & Zhipu AI & undisclosed & API & 2025.4 \\
& Hunyuan & Tencent & undisclosed & API & 2023.9 \\
& DeepSeek-Chat & DeepSeek AI & 236B (MoE) & API & 2024.5 \\
& DeepSeek-V3-0324 & DeepSeek AI & 671B (MoE) & API & 2025.3 \\
& DBRX-Instruct & Databricks & 132B (MoE) & API & 2024.3 \\
& Qwen2-72B-Instruct & Alibaba Cloud & 72B & API & 2024.6 \\
& Qwen2.5-7B-Instruct & Alibaba Cloud & 7B & API & 2024.9 \\
& Qwen3-0.6B & Alibaba Cloud & 0.6B & API & 2025.4 \\
& Qwen3-235B-A22b & Alibaba Cloud & 235B(MOE) & API & 2025.4 \\
& LLaMA2-70B & Meta & 70B & API & 2023.7 \\
& LLaMA3.1 405B & Meta & 405B & API & 2024.7 \\
& Baichuan2-13B-Chat & Baichuan Inc. & 13B & Weights & 2023.12 \\
& Falcon-7B & TII & 7B & Weights & 2023.5 \\
& ChatGLM3-6B & Zhipu AI & 6B & Weights & 2023.1 \\
\midrule
\multirow{13}{*}{\centering Closed-Source} & GPT-4o & OpenAI & undisclosed & API & 2024.5 \\
& GPT-4o-mini & OpenAI & undisclosed & API & 2024.7 \\
& GPT-4 & OpenAI & undisclosed & API & 2023.3 \\
& Gemini 1.5 Pro & Google & undisclosed & API & 2024.5 \\
& Gemini 1.5 Flash & Google & undisclosed & API & 2024.5 \\
& \makecell[l]{Gemini2.5 Pro\\Experimental 03-25} & Google & undisclosed & API & 2025.3 \\
& Claude 3.5 Sonnet & Anthropic & undisclosed & API & 2024.1 \\
& Claude 3.7 Sonnet & Anthropic & undisclosed & API & 2025.2 \\
& GLM-4 & Zhipu AI & undisclosed & API & 2024.1 \\
& Grok 3 & xAI & undisclosed & API & 2025.2 \\
& Doubao-1.5-Pro-256k & ByteDance & undisclosed & API & 2025.1 \\
& Doubao-1.5-Pro-32k & ByteDance & undisclosed & API & 2025.1 \\
& Qwen-Max & Alibaba Cloud & undisclosed & API & 2025.1 \\
\midrule
\multirow{2}{*}{\centering Financial} & Finma-7b-Full & Finma & 7B & Weights & 2023.9 \\
& CFGPT2-7B & TongjiFinLab & 7B & Weights & 2024.8 \\
\bottomrule
\end{tabular}
\caption{Models evaluated in this paper. The "Access" column shows whether we have full access to the model weights or we can only access through API. The “Version Date” column shows the release date of the corresponding version of the model we evaluated.}
\label{tab:llm_summary_merged}
\end{table*}
\end{onecolumn}

\begin{onecolumn} 
\centering
\section{prompt}
\begin{figure*}[htb]
    \centering
    \includegraphics[width=1\linewidth]{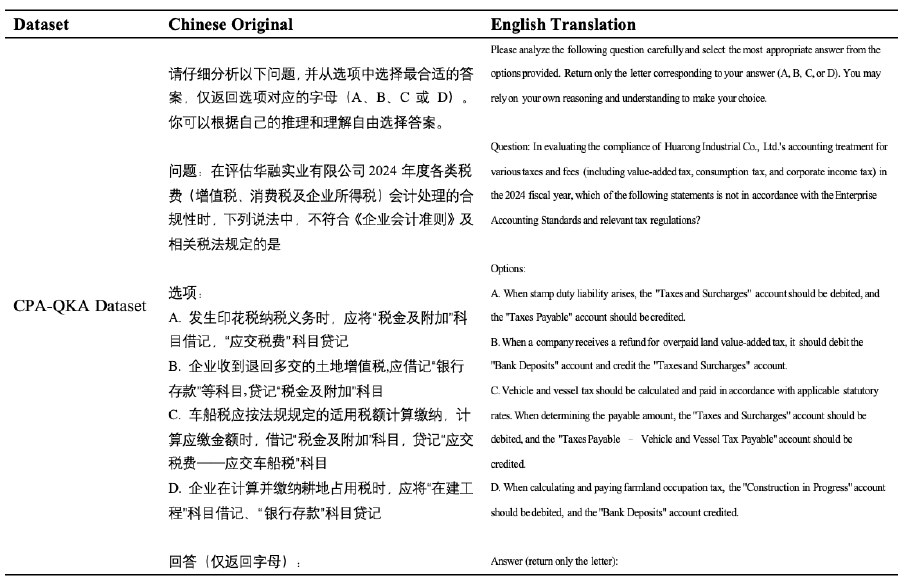}
    \caption{Prompt for multiple-choice questions in Intermediate Financial Accounting. For better readability, the English translation is displayed to the right of the corresponding Chinese text.}
    \label{fig:placeholder}
\end{figure*}
\end{onecolumn}
\end{document}